\documentstyle[12pt]{article}
\newcommand{\be}{\begin{equation}}
\newcommand{\ee}{\end{equation}}
\newcommand{\bea}{\begin{eqnarray}}
\newcommand{\eea}{\end{eqnarray}}
\newcommand{\ba}{\begin{array}}
\newcommand{\ea}{\end{array}}
\newcommand{\beas}{\begin{eqnarray*}}
\newcommand{\eeas}{\end{eqnarray*}}
\newcommand{\bes}{\begin{equation*}}
\newcommand{\ees}{\end{equation*}}

\begin{document}

\title{\bf Holograghic Brownian motion in three dimensional G\"{o}del black hole }
\author{J. Sadeghi \thanks{Email:pouriya@ipm.ir}\hspace{1mm},F. Pourasadollah \thanks
{Email:F. pourasadollah@gmail.com} and H.Vaez \thanks{Email:H. Vaez@umz.ac.ir}\\
{\small {\em  Sciences Faculty, Department of Physics, Mazandaran University,}}\\
{\small {\em P.O.Box 47416-95447, Babolsar, Iran}}}
 \maketitle
\begin{abstract}
By using the AdS/CFT correspondence and G\"{o}del black hole
background, we study the dynamics of  heavy quark under a rotating
plasma. In that case we follow Atmaja (JHEP 1304, 021, (2013)) about
Brownian motion in BTZ black hole. In this paper we receive some new
results for the case of $\alpha^{2}l^{2}\neq1$. This case, we must
redefine the angular velocity
 of string fluctuation. We obtain the time evolution of
 displacement square that angular velocity and show that it behaves as
 a Brownian particle in non-relativistic limit. In this plasma, it seems that relating the Brownian motion
 with physical observables is rather a difficult work. But our results match with Atmaja work in the limit $\alpha^{2}l^{2}\rightarrow1$.
\end{abstract}
\section{Introduction}
  In the last several years, the holographic AdS/CFT \cite{P1}-\cite{P4} has been exploited to study of strongly coupled systems,
   in particular quark gluon plasmas \cite{P5}-\cite{P7}. The quark gluon plasma (QGP) is produced, when two heavy ions
   collide with each other at very high temperature.  A relatively heavy particle, e,g. a heavy quark,
    immerses in a soup of quarks and gluons with small fluctuations
 due to its interaction with constituent of QGP. The random motion of this particle
 well-known
 as Brownian motion \cite{P8,P9}.
The Brownian motion is a universal phenomenon in finite temperature
systems and any particle immersed in a fluid at finite temperature
undergoes Brownian motion. The Brownian motion opens a wide view
from microscopic  nature and studying it by using the AdS/CFT
correspondence. It offers a better understanding of the microscopic
  origin of thermodynamics of black holes, Therefore, it is a natural step to study Brownian motion using the
AdS/CFT correspondence. Particulary, the AdS/CFT correspondence can
be utilized to investigate the Brownian motion for a quark in the
quark gluon plasma.

 In the field theory or boundary side of AdS/CFT story, a mathematical description of Brownian motion is given by the
 Langevin equation which phenomenologically describe the force acting on Bownain particles
 \cite{P9}-\cite{P11} which is given by,
     \begin{equation}
\dot{p}(t)=-\gamma_{0}p(t)+R(t)\,.
     \end{equation}
     where $p$ is momentum of Browniam particle and $\gamma_{0}$ is the friction coefficient.
      These forces originate from losing energy to medium due to friction term (first term) and getting a
       random kick from the thermal bath(second term). One can learn about the microscopic interaction
       between the Brownian particle and the fluid constituents, if  measures these forces be clear.
      By assuming $<m\dot{x}^{2}>=T$, the time evolution of displacement square is given as follows
      [9],
      \begin{equation}\label{eq 2}
        <s(t)^{2}>=<[x(t)-x(0)]^{2}>\approx\ \left\{
                                                 \begin{array}{ll}
                                                  \frac{ T}{m}t^{2}  & \hbox{$(t\ll \frac{1}{\gamma_{0}})$} \\
                                                   2Dt   & \hbox{$(t\gg \frac{1}{\gamma_{0}})$}
                                                 \end{array}
                                               \right.,
      \end{equation}
where $D=\frac{T}{\gamma_{0}m}$ is diffusion constant , T is the
tempreture and $m$ is the mass of Brownian particle.
 At early time, $t\ll \frac{1}{\gamma_{0}}$ (balistic regime), the Brownian particle moves with constant velocity
 $\dot{x}\sim \sqrt{\frac{T}{m}}$, while at the late time, $t\gg \frac{1}{\gamma_{0}}$ (diffusive regime),
 the particle undergoes a random walk.\\
In the gravity or bulk side of AdS/CFT version for Brownian motion,
we need a gravitational analog of a quark immersed in QGP.
 This is acheived by introducing a bulk fundamental string stretching between the boundary at infinity
 and event horizon of an asymptotically AdS black hole background \cite{P12}-\cite{P16}. The dual statement of a quark in QGP
 on the boundary corresponds to the black hole environment that excites the modes of string. In the context
  of this duality, the end of string at the boundary corresponds to the quark which shows Brownian motion and
  its dynamics is formulated by Langevin equation.
    In the formulation of AdS/CFT correspondence, fields of gravitational theory would be related to
     the corresponding boundary theory operators \cite{P3,P4}, such as their boundary value should couple to the operators
      in a way consist with symmetries of problem. In this way, instead of using the boundary field theory to
       obtain the correlation function of quantum operators, we can determine these correlators by the thermal
       physics of black holes and use them to compute the correlation functions. In \cite{P13}-\cite{P16}, the Brownian motion
       has been studied in holographic setting and the time evolution of displacement square.
       If we consider different gravity theories, we know that
       strings will be live in a black hole background and
       excitation  of the modes are done by Hawking radiation of
       black hole. So, different theories of gravity can be
       associated with various plasma in the boundary.
In this paper we follow different works to investigate Brownian
motion of a particle in rotating plasmas. We try to consider the
motion of a particle in two dimensional rotating plasma which its
gravity dual is described by three dimensional G\"{o}del metric
background. In this case,
 we will see that for the $\alpha$ parameter in the G\"{o}del metric, new conditions for Brownian motion will
  be provided. As we know, the Brownian motion of a particle in two dimensional rotating plasma with the corresponding
  gravity of BTZ black hole has been studied in \cite{P15}. The G\"{o}del metric
  background in the special case receives to the BTZ black hole \cite{P17}, so the comparison of our result with what
  obtained in the \cite{P15}, gives us motivation to understanding about the Brownian motion in rotating
  plasmas in general form of background as G\"{o}del black hole.\\
     This paper is arranged as follows. In section 2, we give some review of three dimensional
     G\"{o}del black hole and derive string action from this metric background. Section 3 is devoted to look for
     a holographic realization of Brownian motion  and obtain the solution
      for equation of motion of string in  G\"{o}del black hole geometry.
     We study the Hawking radiation of the transverse modes near the outer horizon of G\"{o}del
  black hole to describe the random motion of the external quark in section 4. In section 5, we make some comments about our
  results and close with conclusions.
\section{Background and String action}
\subsection{G\"{o}del black hole}
Three dimensional G\"{o}del spacetime  is an exact solution of
Einstein-Maxwell theory with a negative cosmological constant and a
Chern-Simons term \cite{P18}.
 When the electromagnetic field acquire a topological mass $\alpha$ Maxwell equation will be modified
  by an addition term. In that case, we receive to Einstein-Maxwell-Chern-Simons system, and the geometry
  is the G\"{o}del space time \cite{P19}. This theory can be viewed as a lower dimensional toy model for the bosonic part of
 five dimensional supergravity theory, so it can be advantage in development of string theory.
 Three dimensional G\"{o}del black holes are alike their higher dimensional counterparts in special properties .
The action of Einstein-Maxwell-Chern-Simons theory in three
dimensions is given by \cite{P20},
\begin{equation}
I=\frac{1}{16\pi G}\int d^{3}x
[\sqrt{-g}(R+\frac{2}{l^{2}}-\frac{1}{4}F_{\mu\nu}
F^{\mu\nu})-\frac{\alpha}{2}\epsilon^{\mu\nu\rho}A_{\mu}F_{\nu\rho}].
\end{equation}
A general spherically symmetric static solution to the above action
in various cases for the $\alpha$ parameter, can be written by
\cite{P21},
\begin{equation}\label{eq 4}
 ds^{2}=\frac{dr^{2}}{h^{2}-pq}+ p dt^{2}+ 2h dt d\phi+ q d\phi^{2},
\end{equation}
where $p$, $q$ and $h$ are functions of $r$ as,
\begin{eqnarray}
p(r)&=&8G\mu,\nonumber\\
q(r)&=&\frac{-4GJ}{\alpha}+2r-2\frac{\gamma^{2}}{l^{2}}r^{2},\nonumber\\
h(r)&=&-2\alpha r,
\end{eqnarray}
with
\begin{equation}
\gamma=\sqrt{\frac{1-\alpha^{2}l^{2}}{8G\mu}}\,\,.
\end{equation}
 The gauge potential is given by,
 \begin{equation}
A=A_{t}(r)dt+A_{\phi}(r)d\phi\,,
\end{equation}
with
\begin{equation}
A_{t}(r)=\frac{\alpha^{2}l^{2}-1}{\gamma\alpha
l}+\varepsilon\qquad,\qquad
A_{\varphi}(r)=\frac{-4GQ}{\alpha}+2\frac{\gamma}{l}r\,\,.
\end{equation}
The parameters $\mu$ and $J $, are mass and angular momentum. The
arbitrary constant $\varepsilon$, is a pure guage. We can rewrite
metric (4) in the ADM form as follows,
\begin{equation}
 ds^{2}=-\frac{\Delta}{q}dt^{2}+\frac{dr^{2}}{\Delta}+ q(
 d\phi+\frac{h}{q}dt)^{2},
\end{equation}
where
\begin{equation}
  \Delta=
  h^{2}-pq=\lambda(r-r_{+})(r-r_{-})\quad\qquad,\qquad\quad\lambda=\frac{2(1+\alpha^{2}l^{2})}{l^{2}},
\end{equation}
and
\begin{equation}
r_{\pm}=\frac{8G}{\lambda}\left[\,\mu\pm\sqrt{\mu^{2}-\frac{\mu
J\lambda}{2\alpha}}\,\right]\,.
\end{equation}
The Hawking temperature that gives the temperature of the plasma, is
\cite{P22}
\begin{equation}
T_{H}=\frac{\lambda(r_{+}-r_{-})}{8\pi\alpha r_{+}}\,.
\end{equation}
 Here $r_{-}$  is the inner horizon and $r_{+}$ is the outer
horizon. In the sector $\alpha^{2}l^{2}>1$, we have real solution
only for $\mu$ negative. In this regime, there  are G\"{o}del
particles and theory supports time-like constants fields. When
$\alpha^{2}l^{2}<1$, $\mu$ has positive values.
  This is the case that black hole will be constructed and theory supports space-like constants fields.
  For $\alpha^{2}l^{2}=1$ , the metric (4) reduces to BTZ metric as
  can be explicitly seen by transforming to the standard frame that is non-rotating at infinity
  with respect to anti-de Sitter space,
\begin{equation}
\phi\rightarrow\phi+\alpha t\qquad,\qquad
r\rightarrow\frac{r^{2}}{2}+\frac{2GJ}{\alpha}\,.
\end{equation}
 In the standard frame, energy and angular momentum
become $M=\mu-\alpha J$ and $J$, instead of $\mu$ and $J$ in
rotating frame.
\subsection{String action}
 In the general case for a $d+2$ dimensional black hole metric background
 is,
\begin{equation}\
 ds^{2}=g_{\mu\nu}(x)dx^{\mu}dx^{\nu}+G_{IJ}(x)dx^{I}dx^{J}.
\end{equation}
Here $x^{\mu}= r,t$ stands for the string worldsheet coordinates and
$X^{I}=X^{I}(x)$ (I,J=1,...,d-2) for the spacetime coordinates. If
we stretch a string along the $r$
 direction and consider small fluctuation in the transverse direction $X^{I}$,
 the dynamics of this string follows from the Nambu-Goto action
 \cite{P13},
\begin{equation}\
 S_{NG}=-\frac{1}{2\pi \acute{\alpha}}\int dx^{2}
 \sqrt{(\dot{X}X^{\prime})^{2}-\dot{X}^{2}X^{\prime 2}}\,.
\end{equation}
If the scalars $X^{I}$ do not fluctuate too far from their equilibrium values ($X^{I}=0$),
 we can expand the above action up to quadratic order in $X^{I}$,
\begin{equation}\
 S_{NG}\approx-\frac{1}{4\pi\acute{\alpha}}\int dx^{2}\sqrt{-g(x)}g^{\mu\nu}G_{IJ}
 \frac{\partial X^{I}}{\partial x^{\mu}}\frac{\partial X^{J}}{\partial
 x^{\nu}}\,.
\end{equation}
In fact, this quadratic fluctuation lagrangian can be interpreted as taking the non-relativistic limit,
 so we must use the dual Langevin dynamics on boundary in the non-relativistic case.
\section{Strings in G\"{o}del black hole}
As we said in the introduction, an external quark is dual to an open string that extends from
 the boundary to the horizon of the black hole \cite{P23}. We can obtain the dynamics of this string in a
 three dimensional G\"{o}del black with the metric background (9) by the  Nambu-Goto action (15)
  in the following form,
\begin{equation}\label{eq 17}
 S_{NG}=-\frac{1}{2\pi\acute{\alpha}}\int dx^{2} \sqrt{\frac{r^{2}}{q
   (r^{2})}+\Delta(r^{2})\phi'^{2}-\frac{q(r^{2})r^{2}}{\Delta (r^{2})}(\dot{\phi}
 +\alpha+\frac{h(r^{2})}{q(r^{2})})^{2}}\,.
\end{equation}
We have obtained the above relation in the standard frame.
   The equation of motion for $\phi$ derived from (\ref{eq 17}) is
\begin{equation}\label{eq 18}
 -\frac{\partial}{\partial t}\left[\frac{r^{2}q}{\Delta\sqrt{-g}}(\dot{\phi}+\alpha+\frac{h}{q})\right]
 +\frac{\partial}{\partial
 r}\left[\frac{\Delta\phi'}{\sqrt{-g}}\right]=0\,.
\end{equation}
\subsection{Trivial solution}
The Nambu-Goto action up to quadratic terms after subsisting the
small fluctuation
 $\phi\rightarrow C+\phi$ under  G\"{o}del metric background in the standard frame is given by
\begin{equation}\
  S_{NG}^{(2)}=-\frac{1}{4\pi\acute{\alpha}}\int dt dr\frac{\Delta^{\frac{3}{2}}\phi'^{2}}{r^{3}
  \left[\frac{\Delta}{q}-q(\frac{h}{q}+
  \alpha)^{2}\right]^{\frac{1}{2}}}-\frac{\Delta^{\frac{1}{2}}\dot{\phi}^{2}}{r\left[\frac{\Delta}{q}-
  q(\frac{h}{q}+
  \alpha)^{2}\right]^{\frac{3}{2}}}\,.
\end{equation}
 By changing coordinate to $s=x-x_{+}$ (where
 $x=\frac{r^{2}}{2}+\frac{2GJ}{\alpha}$) and defining
$\phi(t,s)=e^{-i\omega t}f_{\omega}(s)$, one can write the equation
of motion as following,
\begin{equation}
 W(s)Z(s)\partial^{2}_{s}f_{\omega}+\frac{1}{2}\left[3\partial Z(s)W(s)-\partial W(s)Z(s)\right]
 \partial_{s}f_{\omega}+\omega^{2}f_{\omega}=0\,,
\end{equation}
where
\begin{equation}
 W(s)= \frac{\Delta(s)}{q(s)}-q(s)(\frac{h(s)}{q(s)}+
  \alpha)^{2}\qquad ,\qquad Z(s)=\Delta(s)\,.
\end{equation}
The solution for this equation of motion is the trivial solution for
the relation (\ref{eq 18}). In general, solution of this equation is
very complicated. However for the extremal case $
\mu=\frac{J(1+\alpha^{2}l^{2})}{\alpha l^{2}}$, we can find an
analytical solution as,
\begin{equation}
 f^{\pm}_{\omega}=s^{-1\pm D}Y(s)\qquad,\quad where
\quad D=\sqrt{1+\frac{\omega^{2}l^{2}}{16GM}}\quad with\quad
M=\frac{\mu}{1+\alpha^{2}l^{2}}\,,
\end{equation}
where $Y(s)$ is obtained from the following relation
\begin{eqnarray}
&s \left( c{s}^{2}+4\,z \left( s-\frac{1}{4}\,{\frac
{z}{{\alpha}^{2}}} \right)  \right) {\frac
{d^{2}}{d{s}^{2}}}Y\left(s
\right)+\left(2\,\chi\,\left(c{s}^{2}+4\,z \left(
s-\frac{1}{4}\,{\frac{z}{{ \alpha}^{2}}} \right)  \right)
+2\,c{s}^{2}+10\,sz-3\,{\frac {{z}^{2}}
{{\alpha}^{2}}} \right) {\frac {d}{ds}}Y \left( s \right)\nonumber\\
&+\chi\,
 \left(  \left( \chi-1 \right)  \left( cs+4\,z \right) +2\,cs+10\,z
 \right) Y \left( s \right) =0\,.
 \end{eqnarray}
 where $\chi=-1\pm D$ and $z=\frac{2 x_{+}}{l^{2}}$. The complete solution of Y(s) is derived in the
 Appendix.
With similar argument as in the ref \cite{P15}, this solution is not
acceptable, because it dosn't have
 oscillatory  modes in radial coordinates and also for this trivial constant solution, one can investigate that the
 square root determinant of the worldsheet metric is not real everywhere, therefore it is not physical
 solution. Due to nonphysical motivation about the mentioned
 solution,  we have to consider other approach, which is linear
 solution.
\subsection{Linear solution}
We can take the linear ansatz for the small fluctuation in the
transverse direction $\phi$ to achieve
 a non-trivial solution, so we expand it as,
    \begin{equation}\label{phi}
    \phi(t,r)=wt+\eta(r)\,,
   \end{equation}
where $w$ is a constant angular velocity. By replacing this relation
into equation (\ref{eq 18}), the solution for $\eta$ is obtained as
following,
\begin{equation}
 \eta'(r)=-\frac{\pi_{\phi}}{\Delta}\sqrt{\frac{\frac{r^{2}}{q}
 (\Delta-q^{2}(\frac{h}{q}+\alpha+w)^{2}}{\Delta-\pi_{\phi}^{2}}}\,,
\end{equation}
where $\pi_{\phi}$ is a constant which has a concept as the total
force to keep string moving with linear angular velocity $w$, and
also is related to momentum conjugate of $\phi$ in r direction. At
$r=r_{NH}$, the numerator becomes zero, so the denominator should
also vanish there, because the string solution (\ref{phi}) must be
real everywhere along the worldsheet. For $w=0$ and
$\alpha^{2}l^{2}\neq1$, $r_{NH}$ is given by
\begin{equation}
  r_{NH}^{2}=-\frac{l^{2}}{\gamma^{2}}\left[1\pm\sqrt{1+\frac{8G\gamma^{2}}{l^{2}\alpha^{2}}
  (2\mu-J\alpha)}\right]-\frac{4GJ}{\alpha}\,.
\end{equation}
When $w\neq0$ and $\alpha^{2}l^{2}=1$ , we receive the excepted
relation for BTZ black hole [15]. For $w\neq0$ and
$\alpha^{2}l^{2}\neq1$, we obtain
\begin{equation}
  r_{NH}^{2}=\frac{l^{2}}{\gamma^{2}(\alpha+w)}\left[(w-\alpha)\pm\sqrt{(w-\alpha)^{2}+
  \frac{16G\gamma^{2}}{l^{2}}
  (\mu-\frac{J(\alpha+w)^{2})}{2\alpha}}\right]-\frac{4GJ}{\alpha}\,.
\end{equation}
 According to \cite{P15} we set dominator to zero, so we have,
\begin{equation}
 \pi_{\phi}^{2}=\Delta=(\frac{h\alpha+p}{\alpha})^{2}\qquad,\qquad
 h=h(r_{NH}^{2}|_{w=0})\,.
\end{equation}
The external force $F_{ext}$, can be obtained by considering the
rotation and the topological mass of black hole which is given by,
\begin{equation}
 F_{ext}=\frac{\pi_{\phi}}{2\pi\acute{\alpha}}=\frac{h\alpha+p}{2\pi\acute{\alpha}\alpha}\,.
\end{equation}
After extracting this external force, we can derive the friction
coefficient $\gamma_{0}$ for non-zero $w$, by considering the
relation $p_{\phi}=m_{0}w$ , as
\begin{equation}
\gamma_{0}m_{0}=\frac{q_{NH}}{2\pi\acute{\alpha}}=\frac{r_{NH}^{2}
-\frac{2\gamma^{2}}{l^{2}}(\frac{r_{NH}^{2}}{2}+\frac{2GJ}{\alpha})^{2}}{2\pi\acute{\alpha}}\,.
\end{equation}
With $\alpha ^{2}l^{2}=1$ this coefficient reduce to the excepted
value $\gamma_{0}=\frac{r_{NH}^{2}}{2\pi\acute{\alpha}m_{0}}$ for
BTZ black hole \cite{P15}.\\\\ The Nambu-Goto action, with the small
fluctuation, $\phi\rightarrow wt+\eta(r)+\phi$ under
 the G\"{o}del background becomes,
\begin{equation}
 S_{NG}^{(2)}=-\frac{1}{4\pi\acute{\alpha}}\int dt dr\frac{\Delta^{\frac{3}{2}}\phi'^{2}}{r^{3}
  \left[\frac{\Delta}{q}-q(\frac{h}{q}+
  \alpha+w)^{2}\right]^{\frac{1}{2}}}-\frac{\Delta^{\frac{1}{2}}\dot{\phi}^{2}}{r\left[\frac{\Delta}{q}-
  q(\frac{h}{q}+
  \alpha+w)^{2}\right]^{\frac{3}{2}}}\,.
\end{equation}
The equation of motion from the above Nambu-Goto action is given by,
\begin{equation}\label{eq 32}
  \frac{-r\Delta^{\frac{1}{2}}\ddot{\phi}}{
  \left[\frac{\Delta}{q}-q(\frac{h}{q}+
  \alpha+w)^{2}\right]^{\frac{3}{2}}}+\frac{\partial}{\partial r}\frac{\Delta^{\frac{3}{2}}
  \phi'^{2}}{r\left[\frac{\Delta}{q}-
  q(\frac{h}{q}+ \alpha+w)^{2}\right]^{\frac{1}{2}}}=0\,.
\end{equation}
Solving this equation is quite complicated for more values of $w$.
However, one can find that there are some value like
\begin{equation}\label{eq 33}
w=\frac{x_{-}+(x_{+}-x_{-})\frac{(1-\alpha^{2}l^{2})}{2}}{\alpha
x_{+}l^{2}}=
\frac{r_{-}+(r_{+}-r_{-})\frac{(1-\alpha^{2}l^{2})}{2}}{\alpha
r_{+}l^{2}}\,,
\end{equation}
 where this makes it possible to solve the equation of motion. In derivation the
 r.h side of the above relation we use $r_{-}r_{+}=\frac{4GJ}{\alpha}$.  The special radius $r_{NH}$
  approaches the outer horizon of the G\"{o}del black hole for this value
 of angular velocity $w$, where $\Delta(r_{+})=0$( and $\pi_{\phi}=0$), then the steady state
 solution is the case that $r_{NH}=r_{+}$. From the relation (\ref{eq 33}) for the angular velocity, it is evident that
 we can receive to $w=\frac{r_{-}}{r_{+}}$ for BTZ black hole $(\alpha^{2}l^{2}=1)$. Furthermore for
  $(\alpha^{2}l^{2}>1)$, we can check that $w^{2}<\alpha^{2}$, but there must be some condition on
  $\mu$ and $J$ to have $w^{2}<\alpha^{2}$ for $(\alpha^{2}l^{2}<1)$. We can write the equation of motion
  for this terminal angular velocity with changing coordinate to $s=x-x_{+}$
  as,
\begin{equation}\label{eq 34}
 W(s)Z(s)\phi''_{s}+\frac{1}{2}\left[3\partial Z(s)W(s)-\partial W(s)Z(s)\right]
 \phi'_{s}-\ddot{\phi}_{s}=0\,,
\end{equation}
where
  \begin{eqnarray}
  W(s)=\frac{p}{(2\alpha x_{+})^{2}}\left[s(cs+\lambda \zeta)\right]
  \qquad,\qquad
 Z(s)=\lambda s(s+\zeta),
  \end{eqnarray}
and
\begin{equation}
  c=\lambda-4\alpha^{2}\qquad\qquad,\qquad\qquad\zeta=x_{+}-x_{-}\,.
\end{equation}
As before, we take $\phi(t,s)=e^{-i\omega t}f_{\omega}(s)$, so the
equation (\ref{eq 34}) reduce to,
\begin{equation}\label{eq 37}
 W(s)Z(s)f_{\omega}''+\frac{1}{2}\left[3 Z'(s)W(s)- W'(s)Z(s)\right]
 f_{\omega}'+\omega^{2}f_{\omega}=0\,.
\end{equation}
Consequently, the independent linear solutions to the above
equation, are obtained as bellow
\begin{equation}\label{eq 38}
 f_{\omega}^{\pm}(s)=(\lambda\zeta s)^{\pm i\vartheta} \:_{2}F_{1}(\pm i\vartheta\:
 ,\:\frac{3}{2}\pm i\vartheta\:;\:1\pm2i\vartheta\:,\:\frac{(-\lambda+c)s}{cs+\lambda\zeta})
 (cs+\lambda\zeta)^{\mp i\vartheta}\,,
\end{equation}
where
\begin{equation}
\vartheta=\frac{2\alpha \omega x_{+}}{\lambda\zeta\sqrt{p}}\,,
\end{equation}
or with $\xi=2\zeta=r_{+}^{2}-r_{-}^{2}$ and $
\frac{x_{+}}{\sqrt{p}}=\frac{4GJ}{\alpha r_{-}\sqrt{\lambda}}$, we have
$\vartheta=\frac{16GJ\omega}{\lambda^{\frac{3}{2}}\xi r_{-}}$\,.By
considering the following relation for hypergeometric  functions,
\begin{equation}
_{2}F_{1}(\kappa,\kappa+\frac{3}{2},2\kappa+1,z)=\frac{2^{2\kappa}}
{\kappa+\frac{1}{2}}\left[1+(1-z)^{\frac{1}{2}}\right]^{-2\kappa}\left[\frac{1}{2}+
\kappa(1-z)^{-\frac{1}{2}}\right]\,,
\end{equation}
the equation (\ref{eq 38}) reduces as,
\begin{equation}\label{eq 41}
 f_{w}^{\pm}(s)=\frac{(4\lambda\zeta )^{\pm i\vartheta}}{1\pm2i\vartheta} \frac{\left[1\pm\frac{2i\vartheta}
 {\sqrt{1+\frac{(\lambda-c)s}{cs+\lambda\zeta}}}\right]}
 {\left[1+\sqrt{1+\frac{(\lambda-c)s}{cs+\lambda\zeta}}\right]^{\pm2i\vartheta}}
 (cs+\lambda\zeta)^{\mp i\vartheta} s^{\pm i\vartheta}\,,
\end{equation}
which gives oscillation modes. We have the following asymptotic
behavior from the solutions near the outer horizon $(s\rightarrow
0)$ and the boundary $(s\rightarrow \infty)$,
\begin{equation}\label{eq 41}
f_{w}^{\pm}(s)\sim\ \left\{
 \begin{array}{ll}
 e^{\pm i\omega s_{\star}}  & (s\rightarrow0) \\
\frac{(4\lambda\zeta)^{\pm
i\vartheta}(1\pm\frac{2i\vartheta}{\sqrt{\frac{\lambda}{c}}})}
 {(1\pm2i\vartheta)(1+\sqrt{\frac{\lambda}{c}})^{\pm2i\vartheta}}   & (s\rightarrow\infty)
                                                 \end{array}
                                               \right.,
\end{equation}
with
$s_{\star}=\frac{\vartheta}{\omega}\ln(s)=\frac{16GJ}{\lambda^{\frac{3}{2}}\xi
r_{-}}\ln(s)$.
\section{Displacement Square}
So far, we have succeeded to drive oscillation modes for a string
moving in the G\"{o}del black hole background. In the following, we
follow the same procedure as in \cite{P13,P15} to compute the
displacement square for Brownian motion. In order to achieve this,
we write the solutions for
 bulk equation of motion as a linear combination of $f^{\pm}_{\omega}$ ,
\begin{equation}\label{eq 43}
f_{\omega}(s)=A\left[f_{\omega}^{+}(s)+Bf_{\omega}^{-}(s)\right]e^{-i\omega
t}\,,
\end{equation}
 $A$ and $B$ are constants. By exerting the Neumann boundary
condition near the boundary,
 $\partial_{s}f_{s}(\omega)=0$  with $s=s_{c}\gg0$, to put the UV-cutoff, we
 obtain,
\begin{equation}
B=\frac{(4\lambda\zeta)^{2i\vartheta}(cs_{c}+\lambda\zeta)^{-2i\vartheta}
s_{c}^{2i\vartheta}(1-2i\vartheta)}{\left[1+\sqrt{1+\frac{(\lambda-c)s_{c}}{c
s_{c}+
\lambda\zeta}}\right]^{4i\vartheta}(1+2i\vartheta)}\frac{\left[1+2i\vartheta
\sqrt{1+\frac{(\lambda-c)s_{c}}{c
s_{c}+\lambda\zeta}}\right]}{\left[1-2i\vartheta
\sqrt{1+\frac{(\lambda-c)s_{c}}{c s_{c}+\lambda\zeta}}\right]}
\equiv e^{i\theta_{\omega}}\,,
\end{equation}
Note that the constant B is a pure phase, so by using the equation
 (\ref{eq 41})in the near horizon we can write,
 \begin{equation}\label{eq 45}
\Phi(t,s)=f_{\omega}(s)e^{i\omega t}\sim e^{-i\omega(
t-s_{\star})}+e^{i\theta_{\omega}} e^{-i\omega( t+s_{\star})}\,.
\end{equation}
To regulate the theory, we implement another cutoff near the outer
horizon at $s_{h}=\epsilon ,\epsilon\ll1$ , which called IR-cutoff,
we obtain,
\begin{equation}
B\approx\epsilon^{2i\vartheta}=e^{-2i\vartheta\ln(\frac{1}{\epsilon})}\,.
\end{equation}
If we take $B$ in the terms of $\omega$ by the relation (\ref{eq
43}) only, then $B$ has continues values, since the $\omega$ can
have any value. Using the relation (\ref{eq 45}) for $B$ will
satisfy our requirements to have discrete values in $\epsilon\ll1$.
In this case, the discreteness is \cite{P13,P15},
\begin{equation}
\triangle \vartheta=\frac{\pi}{\ln\frac{1}{\epsilon}}\,,
\end{equation}
where in terms of $\omega$, it is given by
\begin{equation}\label{eq 48}
\triangle \omega=\frac{\lambda^{\frac{3}{2}}\pi
r_{-}\xi}{16GJ\ln\frac{1}{\epsilon}}\,.
\end{equation}
Following the above processes and using IR cutoff to discrete the
continues spectrum , makes it easy to find normalized bases of modes
and to quantize $\phi(t,r)$ by extending in this modes.
\subsection{Brownian particle Location}
In this section we are going to use quantized modes of the string
near the outer horizon of G\"{o}del black hole to describe the
Brownian motion of an external quark. Therefore we
 consider the Nambu-Goto action for certain amount of terminal angular velocity, near the
  outer horizon $(s\rightarrow0)$ ,
\begin{equation}
S_{NG}^{2}\sim\frac{1}{2}\int dt
ds_{\star}(\dot{\Phi}^{2}-\Phi'^{2})\,.
\end{equation}
where
$\Phi\equiv\frac{8GJ}{r_{-}\sqrt{2\pi\lambda\acute{\alpha}}}\phi$\,.
Thus, according to the same procedure for standard scalar fields,
 we introduce the following mode expansions,
\begin{equation}\label{eq 50}
  \phi(t,s)=\sum_{\omega>0}\left[a_{\omega}u_{\omega}(t,s)+
  a_{\omega}^{\dag}u_{\omega}(t,s)^{\ast}\right]\,,
\end{equation}
with
\begin{equation}
u_{\omega}(t,s)=\sqrt{\frac{\lambda^{\frac{3}{2}}\xi r_{-}}
{16GJ\ln(\frac{1}{\epsilon})}}\left[f_{\omega}^{+}(s)+Bf_{\omega}^{-}(s)\right]e^{-i\omega
t}\qquad,\qquad
\left[a_{\omega},a_{\acute{\omega}}^{\dag}\right]=\delta_{\omega\acute{\omega}}\,.
\end{equation}
Now, by considering the above quantum modes on the probe string in
the bulk, we want to work out the dynamics of the endpoint which
corresponds to a external quark. We investigate the wave-functions
of the world-sheet fields in the two interesting regions:
  (i) near the black hole horizon and (ii) close to the boundary.
  From (\ref{eq 41}), near the horizon $(S\sim0)$, the expansion (\ref{eq 50})
  becomes,
\begin{equation}
\phi(t,S\rightarrow0)=\frac{r_{-}^{\frac{3}{2}}\sqrt{2\pi\acute{\alpha}
\lambda^{\frac{5}{2}}\xi}}{(16GJ)^{\frac{3}{2}}\sqrt{\ln\frac{1}{\epsilon}}}
\sum_{\omega=-\infty}^{\infty}\frac{1 }{\sqrt{\omega}}(e^{-i\omega(
t-S_{\star})}+e^{i\theta_{\omega}} e^{-i\omega(
t+S_{\star})})a_{\omega}\,.
\end{equation}
We used $S=2s=r^{2}-r_{+}^{2}$. On the other hand, the expansion
(\ref{eq 50}) at $S=R$ ( the location of the regulated boundary) is
given by,
\begin{equation}
\phi(t,S=R)=\frac{r_{-}^{\frac{3}{2}}\sqrt{2\pi\acute{\alpha}
\lambda^{\frac{5}{2}}\xi}}{(16GJ)^{\frac{3}{2}}\sqrt{\ln\frac{1}{\epsilon}}}
\sum_{\omega>0}\frac{1 }{\sqrt{\omega}}\left[\frac{2^{1+2i\vartheta}
(\lambda \xi
R)^{i\vartheta}(cR+\lambda\xi)^{-i\vartheta}(1-2i\vartheta)}
{\left[1+\sqrt{1+\frac{(\lambda-c)R}{c
R+\lambda\xi}}\right]^{2i\vartheta}\left[1-2i\vartheta
\sqrt{1+\frac{(\lambda-c)R}{c R+\lambda\xi}}\right]} e^{-i\omega t}
a_{\omega}+ h.c\right]\,.
\end{equation}
One can see that there are two modes in the solutions. The outgoing
modes ($\omega>0$) that are excited because of Hawking radiation
\cite{P24,P25} and incoming modes ($\omega<0$) which
 fall into black hole. The outgoing mode correlators are determined by the thermal density matrix,
\begin{equation}\label{eq 54}
  \rho_{0}=\frac{e^{-\beta H}}{Tr(e^{-\beta H})}\qquad,
  \qquad H=\sum_{\omega>0}\omega a_{\omega}^{\dag}a_{\omega}\,,
\end{equation}
 and the expectation value of occupation number is given by the Bose-Einstein distribution,
\begin{equation}
<a_{\omega}^{\dag}a_{\acute{\omega}}>=\frac{\delta\omega\acute{\omega}}
{e^{\beta\omega}-1}\,,
\end{equation}
with $\beta=\frac{1}{T}$. Using the knowledge of relation (\ref{eq
54}) about outgoing modes correlators
 in the bulk, we can investigate the motion of the endpoint
 of the string at $S=R\gg1$. We can also determine the behavior of the Brownian motion,
  by computing displacement square, as came in (\ref{eq 2}). So we can predict the nature of Brownian
   motion of external particle on the boundary. For this purpose, we compute the modes correlators at $S=R\gg1$
   as,
\begin{equation}
<\phi_{R}(t)\phi_{R}(0)>=\sum_{\omega>0}\frac{r_{-}^{3}\pi\acute{\alpha}
\lambda^{\frac{5}{2}}\xi}{(16GJ)^{3}\omega\ln(\frac{1}{\epsilon})}
\frac{[1+4\vartheta(\omega)^{2}]}{\left[1+4\vartheta(\omega)^{2}
(\frac{\lambda(R+\xi)}{c R+\lambda\zeta})\right]}
\left[\frac{2\cos\omega t}{e^{\beta\omega}-1}+e^{-i\omega
t}\right]\,.
\end{equation}
By utilizing (\ref{eq 48}), we can write above relation in the
integral form. We observe that the integral is diverge.
 So we regularize it by normal ordering the $a$
 ,$a^{\dag}$ oscillators $:a_{\omega}a_{\omega}^{\dag}\equiv a_{\omega}^{\dag}a_{\omega}$,
  then we have,
\begin{equation}
<:\phi_{R}(t)\phi_{R}(0):>=\frac{8\lambda
r_{-}^{2}\acute{\alpha}}{(16GJ)^{2}}
\int_{0}^{\infty}\frac{d\omega}{\omega}\frac{(1+4\frac{(16GJ)^{2}\omega^{2}}
{\lambda^{3}\xi^{2} r_{-}^{2}})}{(1+4\frac{(16GJ)^{2}\omega^{2}}
{\lambda^{3}\xi^{2} r_{-}^{2}}(\frac{\lambda(R+\xi)}{c
R+\lambda\xi}))} \left[\frac{2\cos(\omega
t)}{e^{\beta\omega}-1}\right]\,,
\end{equation}
and the displacement square becomes,
\begin{equation}
S_{reg}(t)^{2}\equiv<:[\phi_{R}(t)-\phi_{R}(0)]^{2}:>=\frac{16\lambda
r_{-}^{2}\acute{\alpha}}{(16GJ)^{2}}
\left[\frac{(\lambda-c)R}{\lambda(R+\xi)}I_{1}+\frac{cR+\lambda\xi}{\lambda(R+\xi)}I_{2}\right]\,,
\end{equation}
with
\begin{equation}\label{eq 59}
I_{1}=4\int_{0}^{\infty}\frac{dy}{y(1+a^{2}y^{2})}\frac{\sin^{2}(\frac{ky}{2})}{e^{y}-1}
\qquad,\qquad
I_{2}=4\int_{0}^{\infty}\frac{dy}{y}\frac{\sin^{2}(\frac{ky}{2})}{e^{y}-1}\,,
\end{equation}
and we have defined
\begin{equation}\label{eq 60}
y=\beta\omega,\qquad k=\frac{t}{\beta},\qquad
a^{2}=4\left(\frac{16GJ}{\lambda\xi r_{-}
\beta}\right)^{2}\left(\frac{R+\xi}{cR+\lambda\xi}\right)\,.
\end{equation}
The evaluation of these integrals and their behavior for $R\gg1$ and
$a\gg1$ can be found in Appendix B of \cite{P13}. From the relation
(\ref{eq 60}), we can see that when $R\gg1$, we have $a\propto
\frac{1}{c}$. Thus in general case
 for $a$, we use the following relations for the integrals (\ref{eq
 59}),

\begin{eqnarray}
  &I_{1}=\frac{1}{2}\left[e^{\frac{k}{a}}Ei(-\frac{k}{a})+e^{\frac{-k}{a}}Ei(\frac{k}{a})\right]
  + \frac{1}{2}\left[\psi(1+\frac{1}{2\pi a})+\psi(1-\frac{1}{2\pi a})\right]
  -\frac{\pi}{2}(1-e^{\frac{|k|}{a}})\cot\frac{1}{2a} \nonumber\\
 &+\log(\frac{2a\sinh \pi k}{k})+ \frac{e^{-2\pi|k|}}{2}\left[\frac{_{2}F_{1}
 (1,1+\frac{1}{2\pi a} ,2+\frac{1}{2\pi a};e^{-2\pi|k|})}{1+\frac{1}{2\pi a}}+
   \frac{_{2}F_{1}(1,1-\frac{1}{2\pi a}
, 2-\frac{1}{2\pi a};e^{-2\pi|k|})}{1-\frac{1}{2\pi a}}\right]\,,
\end{eqnarray}
\begin{equation}
  I_{2}=\log(\frac{\sinh\pi k}{\pi k})\,.
\end{equation}
However, for $R\gg1$ and $c\ll1$  $(\alpha^{2}l^{2}\rightarrow1)$,
then $a\gg1$, one can utilize the following relation for $I_{1}$ and
$I_{2}$,

\begin{equation}
                                  I_{1}=\left\{
                                                \begin{array}{ll}
                                                  \frac{\pi k^{2}}{2a}+O(a^{-2}) \\
                                                   \pi k + O(\log k)
                                                 \end{array}
                                                    \right.\qquad\qquad                   I_{2}=\left\{
                                                \begin{array}{ll}
                                                  O(a^{0}) & \hbox{$(k\ll a)$}\\
                                                   \pi k + O(\log k)& \hbox{$(k\gg a)$}
                                                 \end{array}
                                                    \right.\,.
 \end{equation}
 Therefore, $S_{reg}(t)^{2}$ has the following form
 \begin{equation}\label{eq 64}
        <S_{reg}(t)^{2}>= \left\{
                                    \begin{array}{ll}
                                       \frac{ 16r_{-}^{3}\lambda\pi\acute{\alpha}}{(16GJ)^{3}\beta}\left[
                                       \frac{\xi(\lambda-c)(cR+\lambda\xi)^{\frac{1}{2}}}{4(R+\xi)^{\frac{1}{2}}}
                                      \right]t^{2}+O(\frac{cR+\lambda\xi}{R+\xi})   & \hbox{$(t\ll \beta)$} \\
                                        \frac{16r_{-}^{2}\lambda\pi\acute{\alpha}}{(16GJ)^{2}\beta}t+
                                        O(\log\frac{t}{\beta})                  & \hbox{$(t\gg \beta)$}
                                       \end{array}
                                               \right.,
      \end{equation}
One can check that the displacement square (\ref{eq 64}) is
consistent with BTZ black hole in ref.\cite{P15}
 by setting $c=0$ or $\alpha^{2}l^{2}=1$. In that case, the $w$ vanishes for $J=0$ (or $r_{-}=0$), but
when $\alpha^{2}l^{2}\neq1$ , the $w$ will have zero value only for
$\frac{J}{\alpha l^{2}}=(\alpha^{2}l^{2}-1)\mu$ (see relation
\ref{eq 33}). Then our static solution is achieved by this
condition.
 The diffusion constant from  (\ref{eq 64}) is given by,
 \begin{equation}
 D=\frac{\lambda \pi \acute{\alpha}\alpha^{2}}{2 r_{+}^{2}}T\,.
 \end{equation}
 So, the relaxation time of Brownian
 particle is as following,
 \begin{equation}
 t_{c}=\frac{1}{\gamma_{0}}=\frac{\lambda m_{0}\pi \acute{\alpha}\alpha^{2}}{2
 r_{+}^{2}}\,.
 \end{equation}
The mass of external particle, $m_{0}$, can be computed by using the
total energy and momentum of string \cite{P26} under the metric
background (\ref{eq 4}),
\begin{equation}
E=\frac{1}{2\pi\acute{\alpha}}\int dr \pi_{t}^{0}\qquad,\qquad
p_{\phi}=\frac{1}{2\pi\acute{\alpha}}\int dr \pi_{\phi}^{0}\,,
\end{equation}
with
\begin{equation}
  \pi_{t}^{0}=\frac{\phi'^{2}}{\sqrt{-g}}(g_{t\phi}^{2}-g_{tt}g_{\phi\phi})
 - \frac{g_{rr}}{\sqrt{-g}}(g_{tt}+g_{t\phi}\dot{\phi})\quad,\quad \pi_{\phi}^{0}=
  \frac{g_{rr}}{\sqrt{-g}}(g_{t\phi}+g_{\phi\phi}\dot{\phi})\,.
\end{equation}
Then we have,
\begin{equation}
 E=\frac{\alpha}{2\pi\acute{\alpha}}\int ds\frac{c(s+x_{+})+\lambda x_{+}}
 {\sqrt{\lambda p (s+\zeta)(cs+\lambda\zeta)}}\qquad,\qquad
 p_{\phi}=\frac{1}{2\pi\acute{\alpha}}
 \int ds\frac{-c(s+x_{+})+\lambda x_{-}}
 {\sqrt{\lambda p (s+\zeta)(cs+\lambda\zeta)}}
\end{equation}
One can check that after putting $c=0 $ in the above relation, the
result of integral is as excepted for BTZ black hole. However for
$c\neq0 $ we obtain,
\begin{equation}
 E=\alpha\frac{\sqrt{(cR+\lambda\xi)(R+\xi)}-\sqrt{\lambda\xi^{2}}}{\sqrt{\lambda p}}
 +2\alpha \frac{\lambda+c}{\lambda}\sqrt{\frac{p}{\lambda c}}\ln\left
 [\frac{\sqrt{cR+\lambda\xi}+\sqrt{c(R+\xi)}}
 {\sqrt{c\xi}+\sqrt{\lambda\xi}}
 \right]\,,
\end{equation}

\begin{equation}
 p_{\phi}=-\frac{\sqrt{(cR+\lambda\xi)(R+\xi)}-\sqrt{\lambda\xi^{2}}}{\sqrt{\lambda p}}
 +2\frac{\lambda-c}{\lambda}\sqrt{\frac{p}{\lambda c}}\ln\left
 [\frac{\sqrt{cR+\lambda\xi}+\sqrt{c(R+\xi)}}
 {\sqrt{c\xi}+\sqrt{\lambda\xi}}
 \right]\,,
\end{equation}
where $\sqrt{\frac{p}{\lambda}}=\frac{r_{+}+r_{-}}{2}$. Then the
mass is defined as,
\begin{equation}
m_{0}^{2}=E^{2}-p_{\phi}^{2}\,.
\end{equation}
So, we can see  correspondence between  equation (64) and (72) with
some condition. It may be interesting to compare two equations with
$ \alpha^{2}l^{2}\rightarrow 1$.

\section{Conclusion}
In this paper, we used Ref. \cite{P15} and obtained the time
evolution of the displacement square for realization of the Brownian
motion of an external quark in a plasma which corresponds to
G\"{o}del black hole. We showed to have a acceptable solution with
oscillatory modes, we had to change the definition of  terminal
angular velocity. We found that in general case
$(\alpha^{2}l^{2}\neq1)$ our results for displacement square are
different in the comparison with Ref. \cite{P15}. However, in
$\alpha^{2}l^{2}=1$ limit, We confirmed that our results are agree
with the work of Atmaja \cite{P15}. We derived the physical mass,
but we found that relating the displacement square with physical
observables is a difficult work. This is the problem that we would
like to consider in future work. Also we would like to investigate
the Brownian motion of external quarks in different environments, in
particular plasmas which correspond to metric
backgrounds as Lifshitz geometry \cite{P27} and metric
backgrounds with Hyperscaling violation \cite{P28,P29}.\\\\
 \appendix \textbf{Appendix}:\\
 The solution to the following differential equation
 \begin{eqnarray}
&s \left( c{s}^{2}+4\,z \left( s-\frac{1}{4}\,{\frac
{z}{{\alpha}^{2}}}
 \right)  \right) {\frac {d^{2}}{d{s}^{2}}}Y \left( s \right) +
 \left( 2\,\chi\, \left( c{s}^{2}+4\,z \left( s-\frac{1}{4}\,{\frac {z}{{
\alpha}^{2}}} \right)  \right) +2\,c{s}^{2}+10\,sz-3\,{\frac
{{z}^{2}} {{\alpha}^{2}}} \right) {\frac {d}{ds}}Y \left( s
\right)\nonumber\\ &+\chi\,
 \left(  \left( \chi-1 \right)  \left( cs+4\,z \right) +2\,cs+10\,z
 \right) Y \left( s \right) =0
 \end{eqnarray}
 can be obtained analytically as
  \begin{eqnarray}
&Y \left( s \right) =C_{1}\,{s}^{-3/2-\chi}\sqrt [4]{s \left( cs+4
\,z \right) {\alpha}^{2}-{z}^{2}} \left( {\frac {s}{-c\alpha\,s-2\,z
\alpha+z\sqrt {\lambda}}} \right)
^{\frac{1}{2}(1-\frac{p(\alpha)}{r(\alpha)q(\alpha)})}
\nonumber\\&\left( {\frac {-\alpha\,\sqrt {\lambda}s+2\,s{
\alpha}^{2}-z}{-z\sqrt {\lambda}+\alpha\, \left( cs+2\,z
 \right) }} \right) ^{\frac{1}{2}(1+\frac{3\sqrt{-c^{2}}}{4r(\alpha)q(\alpha)})}
 \left( -c\alpha\,s-2\,z\alpha+z\sqrt {\lambda} \right)\nonumber\\&
_{2}F_{1}\left(E(\alpha)(T(\alpha) + U(\alpha)),E(\alpha)(T(\alpha)
+ U(\alpha)),
(1-\frac{p(\alpha)}{q(\alpha)r(\alpha)});\frac{c\alpha\sqrt{\lambda
s}}{(-z\sqrt{\lambda}+\alpha(cs+2z))r(\alpha)}
  \right)+\nonumber\\&
  C_{2}\,{s}^{-3/2-\chi}\sqrt [4]{s \left( cs+4
\,z \right) {\alpha}^{2}-{z}^{2}} \left( {\frac {s}{-c\alpha\,s-2\,z
\alpha+z\sqrt {\lambda}}} \right)
^{\frac{1}{2}(1+\frac{p(\alpha)}{r(\alpha)q(\alpha)})}
\nonumber\\&\left( {\frac {-\alpha\,\sqrt {\lambda}s+2\,s{
\alpha}^{2}-z}{-z\sqrt {\lambda}+\alpha\, \left( cs+2\,z
 \right) }} \right) ^{\frac{1}{2}(1+\frac{3\sqrt{-c^{2}}}{4r(\alpha)q(\alpha)})}
 \left( -c\alpha\,s-2\,z\alpha+z\sqrt {\lambda} \right)\nonumber\\&
_{2}F_{1}\left(E(\alpha)(T(\alpha) +R(\alpha)),E(\alpha)(T(\alpha) +
R(\alpha)),
(1+\frac{p(\alpha)}{q(\alpha)r(\alpha)});\frac{c\alpha\sqrt{\lambda
s}}{(-z\sqrt{\lambda}+\alpha(cs+2z))r(\alpha)}
  \right)
  \end{eqnarray}
where $ E,R,T,U,r,q,p(\alpha)$ are given by relations (74-78)

  \begin{equation}
E(\alpha)=\frac{1}{ 2\sqrt {-c-8\,{\alpha}^{2}+4\,\alpha\,\sqrt
{\lambda}}
  \left( \alpha+1/2\,\sqrt {\lambda} \right)  \left( 2\,\alpha\,
  \sqrt {\lambda}+\lambda \right) }=\frac{1}{4\sqrt{\lambda} r^{2}(\alpha)q(\alpha)}
  \end{equation}
 \begin{equation}
  r(\alpha)=\sqrt {-c-8\,{\alpha}^{2}+4\,\alpha\,\sqrt {\lambda}}\quad,
q(\alpha)=\left( \alpha+1/2\,\sqrt {\lambda} \right)\quad,
p(\alpha)=\sqrt {- \left( \chi+1 \right) ^{2}{c}^{2}}
\end{equation}

 \begin{equation}
T(\alpha)=\left(  \left( 2\,{\alpha}^{2}+\frac{1}{4}\,c \right)
 \sqrt {\lambda}+\alpha\,  \lambda  \right)
  \sqrt {-c-8\,{\alpha}^{2}+4\,\alpha\,\sqrt {\lambda}}
\end{equation}
 \begin{equation}
U(\alpha)= \left( \frac{3}{2}\,\alpha+\frac{3}{4}\,\sqrt {\lambda}
\right)
  \sqrt {-{c}^{2}  \lambda  }-\sqrt {- \left( \chi+1 \right) ^{2}{c}^{2}}
  \left( 2\,\alpha\,\sqrt {\lambda}+\lambda\right)
\end{equation}
 \begin{equation}
  R(\alpha)=\left( \frac{3}{2}\,\alpha+{\frac {3}{4}}\,\sqrt {\lambda} \right)
     \sqrt {-{c}^{2}  \lambda  }+\,\sqrt {- \left( \chi+1 \right) ^{2}{c}^{2}}
      \left( 2\,\alpha\,\sqrt {\lambda}+\lambda \right)
\end{equation}

\end{document}